\documentclass[aps,prb,twocolumn,nofootinbib,citeautoscript, 10pt]{revtex4-2}  
\usepackage{amsmath,amssymb,bm} 
\usepackage{graphicx}

\usepackage[tight]{subfigure} 

\usepackage{color} 
\usepackage[papersize={8.5in,11in}]{geometry}
\usepackage[colorlinks=true]{hyperref}
\hypersetup{
    bookmarks=true,         
    unicode=false,          
    pdftoolbar=true,        
    pdfmenubar=true,        
    pdffitwindow=false,     
    pdfstartview={FitH},    
    pdfkeywords={keyword1} {key2} {key3}, 
    pdfnewwindow=true,      
    colorlinks=true,       
    linkcolor=magenta, 
    citecolor=blue,        
    filecolor=magenta,      
    urlcolor=blue           
} 

\usepackage{amsfonts,bbold}
\usepackage{upgreek}
\usepackage{slashed}
\usepackage{latexsym}
\usepackage{ulem}

\usepackage{wrapfig} 
\usepackage{wasysym}			
\usepackage{amsthm}
\usepackage{xcolor}
\usepackage{hyperref} 
\usepackage{breakurl}

\def \nn{\nonumber \\}

\newcommand{\be}{\begin{equation}}
\newcommand{\ee}{\end{equation}}

\renewcommand{\vec}[1]{{\bf #1}}

\renewcommand{\epsilon}{\varepsilon}

\begin{document}

\title{Thermopower in an anisotropic two-dimensional Weyl semimetal}
\author{Ipsita Mandal$^{1,2}$}
\author{Kush Saha$^{3,4}$}
\affiliation{$^{1}$Laboratory of Atomic And Solid State Physics, Cornell University, Ithaca, NY 14853, USA\\
$^{2}$Faculty of Science and Technology, University of Stavanger, 4036 Stavanger, Norway\\
$^{3}$National Institute of Science Education and Research, Jatni, Khurda 752050, Odisha, India\\
$^{4}$Homi Bhabha National Institute, Training School Complex, Anushakti Nagar, Mumbai 400094, India}

\begin{abstract}
We investigate the generation of an electric current from a temperature gradient in a two-dimensional Weyl semimetal with anisotropy, in both the presence and absence of a quantizing magnetic field. We show that the anisotropy leads to doping dependences of thermopower and thermal conductivities which are different from those in isotropic Dirac materials. Additionally, we find that a quantizing magnetic field in such systems leads to an interesting magnetic field dependence of the longitudinal thermopower, resulting in unsaturated thermoelectric coefficients. Thus the results presented here will serve as a guide to achieving high thermopower and a thermoelectric figure-of-merit in graphene-based materials, as well as organic conductors such as $\alpha$-(BEDT-TTF)$_2$I$_3$. 
\end{abstract}

\maketitle

\tableofcontents

\section{Introduction}
Since the discovery of Dirac materials in both two and three dimensions, there has been an upsurge in the study of thermopower in these systems, in both the presence and absence of a quantizing magnetic field \cite{Castro2009,Checkelsky2009,Sarma2009,Philip2009,zhu2010,Bergman2010,Tiwari2010,Alexey2011}. This is because the thermopower is a sensitive and powerful tool to probe transport properties, involving different scattering mechanisms in materials. Two-dimensional (2D) graphene and related 2D Dirac materials exhibit anomalous and universal thermoelectric properties due to the Weyl/Dirac dispersion of the emergent quasiparticles \cite{Wei2009,zhu2010}. Similarly, 3D Weyl systems exhibit anomalous thermal properties due to the Berry curvature \cite{girish1,Girish2017,Gegory2014,Gegory2016,GirishTiwari2017,Liang2017,Vozmediano2018}. Moreover, the 3D Dirac and Weyl materials give rise to unsaturated thermopower, which in turn leads to large thermoelectric figure-of-merit in the presence of a quantizing magnetic field \cite{Skinner2018}. 

Despite much work on the transport properties in Dirac/Weyl materials \cite{piet2014,Huang2013,mikito2014,hosur2012,karl2014,kamran2013,prl_niu,Liang2017,kamran2015,Bardarson2017,gorbar2017,trivedi2017}, the thermoelectric properties in the (relatively newly discovered) 2D anisotropic Dirac materials (such as VO$_2$/TiO$_3$ \cite{pardo, pardo2, banerjee}, organic salts \cite{kobayashi,suzumura}, and deformed graphene \cite{hasegawa,orignac,montambaux1,montambaux2}), having a quadratic dispersion in one direction and a linear dispersion along the orthogonal direction, have not been explored so far in detail. This is in part because there is a lack of naturally occurring materials with such anisotropic dispersions and, in part, because the anisotropy leads to complexities in finding the analytical expressions for relevant thermoelectric coefficients involving different scattering mechanisms (compared to the in-plane and out-of-plane anisotropy in double-Weyl materials \cite{Gegory2016}). Due to the anisotropic dispersion, these 2D semi-Dirac materials exhibit unconventional electric and magnetic properties, as opposed to the isotropic Weyl/Dirac systems \cite{landau-level,moon}. Since transport coefficients such as thermal conductivity and thermoelectric coefficients are determined by the bandstructure and scattering mechanisms, it is natural to ask how such anisotropic dispersion can be leveraged against the thermoelectric properties of these 2D systems, both in the presence and absence of a quantizing magnetic field. Specifically, does this anisotropy give rise to interesting field-, temperature-, and doping-dependence of the thermoelectric coefficients? 

To address the points outlined above, we study the electrical and thermal transport in an anisotropic 2D Dirac/Weyl system, both in the absence and in the presence of an external magnetic field applied perpendicular to the plane of the 2D semimetal. We show that the thermopower in the absence of a magnetic field exhibits a complex dependence on the chemical potential and temperature, in contrast to its isotropic counterpart \cite{Sarma2009,Liang2017}. The forms of the transport coefficients of a 2D anisotrpic Dirac/Weyl semimetal differ from those of a 3D double-Weyl case because the density of states (DOS) is different for these systems at a finite chemical potential. Specifically, the DOS of a 3D anisotropic double-Weyl dispersion turns out be $\rho(\epsilon)\sim |\epsilon|$ (where $\epsilon$ denotes the energy of band in question), which simplifies the analytical expressions for the thermoelectric coefficients \cite{Gegory2016}. In contrast, the DOS of a 2D anisotropic Dirac/Weyl dispersion goes as $\rho(\epsilon)\sim \sqrt{|\epsilon|}$ which, in turn, leads to complicated expressions for the thermoelectric transport coefficients, and hence to complex chemical potential and magnetic field-dependent thermopower.

 We also show that the presence of an external magnetic field leads to interesting field-dependent transport properties, including that of an unsaturated thermopower. This field-dependence differs significantly not only from its isotropic counterpart, but also from 3D Dirac/Weyl systems \cite{Skinner2018}. This is attributed to the fact that, for such anisotropic Dirac/Weyl systems, the field-dependence of the Landau level spectrum $  \text{sgn}(n) \left[ \left( n +1/2 \right) H \right ]^{2/3}$ in the regime of a quantizing magnetic field (where $n \in {\mathbb{Z}} $ and $H$ is the applied magnetic field \cite{landau-level}), differs from that of 2D and 3D isotropic Dirac/Weyl systems (with Landau levels given by $ \sim \text{sgn} (n) \sqrt{n\,H}$). We note that a similar anisotropic situation arises in a 3D double-Weyl material \cite{Gegory2016}, where the anisotropy arises due to a linear dispersion along one of the three orthogonal directions and a quadratic dispersion along the plane perpendicular to it. 

The rest of the paper is organized as follows. In Sec.~\ref{model}, we introduce anisotropic 2D Dirac/Weyl model Hamiltonian and define the thermoelectric coefficients.
In Sec.~\ref{free case}, we provide analytical expressions for thermoelectric coefficients in zero magnetic field. We then compare our results with the case of isotropic Dirac dispersion.  
In Sec.~\ref{disorder} and \ref{coulomb}, we present the results for diffusive transport and electron-electron interactions, respectively. We then extend these results to those in presence of quantizing ({\it i.e.} high) magnetic fields in Sec.~\ref{high magnetic field}, and low magnetic fields in Sec.~\ref{weak_field}, and discuss the unsaturated thermopower. Finally, we conclude with a discussion of the possible future directions in Sec.~\ref{conclusion}.

\section{Model and formalism}
\label{model}

 We consider a model of a 2D anisotropic Weyl fermion (AWF), with the Hamiltonian \cite{pardo, pardo2, banerjee} 
\begin{align}
\label{ham-continuum}
H_{\rm AWF} =\frac{\hbar^2\, k_x^2}{2\,m}\,\sigma_x +\hbar \, v\, k_y \,\sigma_y ,
\end{align}
where $\sigma_i$'s are Pauli matrices, $(k_{x},k_{y})$ are the momenta in the $x$ and $y$ directions, respectively, $m$ is the effective mass along the $x$-axis, and $v$ is the effective velocity along the $y$-axis. We will use $a= \frac{\hbar^2 }{2\,m}$ and
$b=\hbar \, v$ in the equations for simplifying the expressions. With these notations, the spectrum of Eq.~(\ref{ham-continuum}) is found to be $\epsilon_{\mathbf k}^\pm = \pm\sqrt{a^2\,k_x^4+b^2\,k_y^2}$. This anisotropic nature of the spectrum is expected to manifest in the thermoelectric properties of the system.

The response matrix, which relates
the resulting generalized currents to the driving forces, can
be expressed in terms of some kinetic coefficients. We will use the relations obtained from
the Boltzmann formalism, such that the response matrix takes the form \cite{mermin}:
\begin{align}
 \left( \begin{array}{c}
\mathbf J_\alpha\\
\mathbf{J}^Q_\alpha 
\end{array} \right) & = 
 \left( \begin{array}{c  c} 
L_{\alpha \beta }^{11} & L_{\alpha \beta }^{12} \\
 L_{\alpha \beta }^{21} & L_{\alpha \beta }^{22}
 \end{array} \right)
\left( \begin{array}{c}
 \mathbf{E}_\beta \\
-\mathbf{\nabla}_\beta T 
\end{array} \right),
\end{align}
where $(\alpha,\beta) \in (x,y) $, $\mathbf J ^Q$ is the heat current and $\mathbf J $ is the electrical current at temperature $T$, in the presence of an electric field $\mathbf E$. For transport along the electric field and temperature gradient, 
the expressions for the longitudinal thermoelectric coefficients are given by \cite{mermin}:
\begin{align}
& L_{\alpha \alpha }^{11} = \sigma _{\alpha \alpha }=  \mathcal{L}_{\alpha }^0\,, \quad
L_{\alpha \alpha }^{21} = T\,  L_{\alpha \alpha }^{12}=\frac{-\mathcal{L}_{\alpha }^1}{e}\,, 
\nn &
L_{\alpha \alpha }^{22} = \frac{\mathcal{L}_{\alpha}^2}{e^2\, T}\,,
\end{align}
with
\begin{align}
\label{eq:transcoeff}
& \mathcal{L}_{\alpha }^n 
\nn &
=- e^2\sum_{s=\pm}\int \frac{d^2 \mathbf{k}}{(2\,\pi )^2}\,  \tau \left(\epsilon^s_{\bf k}\right)
\frac{\partial f (\epsilon ^s _{\bf k}) }{\partial \epsilon ^s _{\bf k}}
\left(\frac{1}{\hbar }\frac{\partial \epsilon^s_{\bf k}}{\partial k_{\alpha}}\right)^2 
  \left(\epsilon_{\bf k}^s-\mu \right)^n,
\end{align}
where $s=\pm$ is the band index, $e$ is the electric charge, $\mu$ is the chemical potential and $f(\epsilon)= \frac{1}{1+e^{\beta \,(\epsilon -\mu )}}$ is the Fermi-Dirac distribution at inverse temperature $\beta=\frac{1}{k_B\,T}$ ($k_B$ is the Boltzmann constant). The thermal conductivity and the Seebeck coefficient can now be defined as:
\begin{align}
\label{eq:kappa}
\kappa _{\alpha  \alpha } & =L_{\alpha  \alpha }^{22}-L_{\alpha  \alpha }^{21}
\left(L^{11}\right)^{-1}_{\alpha  \alpha }
\,L_{\alpha  \alpha }^{12}
\quad \text{and} \quad 
S_{\alpha\alpha } =\frac{L_{\alpha  \alpha }^{12}}{L_{\alpha  \alpha }^{11}}\,,
\end{align}
respectively. The Seebeck coefficient describes the voltage generation due to
a temperature gradient. In the presence of transverse currents, $\kappa$ and $S$ can be written in a more general form \cite{mermin}:   
\begin{align}
\label{eq:general}
\kappa _{\alpha  \beta } & =L_{\alpha  \beta }^{22}
-\sum \limits _{\gamma, \rho} L_{\alpha  \gamma }^{21}\left( L ^{11}\right)^{-1}_{\gamma  \rho}
\, L_{\rho  \beta }^{12}\,,
\quad \nonumber\\ \quad 
S_{\alpha\beta } &= \sum \limits _{\gamma}  (L^{11})^{-1} _{\alpha  \gamma } \,L_{\gamma  \beta }^{12}\,.
\end{align}
The diagonal elements of the matrix $S$ are called Seebeck coefficients (or thermopower), and the off-diagonal components are termed as Nernst coefficients.

Let us denote an external magnetic field by $\mathbf H$ with magnitude $H$. 
In the following sections, we will mainly focus on the thermopower $S_{\alpha\alpha}$, in both the absence (when $H=0$) and the presence (when $H\ne 0$) of transverse thermoelectric coefficients $L_{xy}$.

For the anisotropic dispersion in Eq.~(\ref{ham-continuum}), we follow the methods outlined in Ref.~\cite{park}.
With the parametrization $ k_x= \text{sign}[\cos\theta] \left( \frac{r \,|\cos\theta|} {a} \right )^{1/2}$ and $ k_y= \frac{ r \, \sin\theta}{b}$ with $r \geq 0$, the energy eigenvalues take the simple form $\epsilon_{\bf k}^\pm = \pm r$. The Jacobian of this transformation is given by:
\begin{align}
\label{eq:jacobian}
\mathcal {J}(r,\theta) =&
\begin{vmatrix}
 \frac{\partial k_x}{\partial r}& \frac{\partial k_x}{\partial \theta }\\
\frac{\partial k_y }{\partial r}& \frac{\partial k_y}{\partial \theta }
\end{vmatrix}
 =
\begin{vmatrix}
 \frac{1} {2} \left( \frac{ |\cos\theta| } {a\, r} \right )^{1/2} & -\frac{\sin \theta} {2}   \left( \frac{r } {a \, |\cos\theta|} \right )^{1/2} \\
\sin \theta & r \cos \theta 
\end{vmatrix}
\nn & =\sqrt{\frac{r} {4\,a\, b^2 \, |\cos\theta|}} \,.
\end{align} 
Let us apply this convenient parametrization for calculating the DOS at energy $\epsilon >0$, which is given by:
\begin{align}
\label{eq:rho}
\rho (\epsilon) & = \int \frac{d^2\vec k}{ \left( 2\, \pi \right)^2} \,\delta \left( \epsilon -\epsilon_{\bf k}^+\right)
\nn & = \int_0^\infty dr \int_0^{2\,\pi}  \frac{ d\theta }{ \left( 2\, \pi \right)^2} \, \mathcal {J}(r,\theta)\,\delta \left( \epsilon - r\right)
\nn & =  \int_0^{2\,\pi} \frac{ d\theta }{ \left( 2\, \pi \right)^2} \sqrt{\frac{\epsilon } {4\,a\,  b^2 \, |\cos\theta|}}
=\frac{ 10.4882} {8\,\pi^2 } \sqrt{\frac{\epsilon } {a\,  b^2}}\,.
\end{align}
Clearly, the DOS of the AWF differs from its isotropic counterpart, i.e, graphene, where $\rho(\epsilon)\sim |\epsilon|$ (see Appendix~\ref{dirac}). Thus it is expected to have different thermopower and thermal conductivities, depending on the scattering mechanisms. However, it is not obvious how strongly this anisotropy will manifest in the thermoelectric coefficients as a function of  $\mu$ and $T$. In the following sections, we therefore compute the thermoelectric coefficients \rm {(i)} for the free Hamiltonian, \rm{(ii)} in the presence of short-range disorder, and \rm{(iii)} in the presence of charge impurities. We then compare the results with those obtained for graphene. We also compare the thermoelectric coefficients with those of the isotropic and anisotropic 3D Dirac materials, wherever deemed necessary. Finally, we consider the case where an external magnetic field is applied, in order to determine the power-law dependence of the thermoelectric coefficients on the applied field strength.

\section{Thermoelectric response for the free Hamiltonian}
\label{free case}
Using the semiclassical approach for calculating the dc conductivity by assuming an energy and momentum independent scattering time $\tau$, we get:
\begin{widetext}
\begin{align}\label{sigma}
\sigma _{x x}^{\text{dc}} 
=  \mathcal{L}_{x }^0   & = 
 \frac{e^2\, \tau\,\sqrt{a}\,\beta} {8\,\pi^2\,\hbar^2\, b}
\int_0^\infty dr \int_0^{2\,\pi} d\theta  \,r^{3/2} \,|\cos \theta|^{5/2} 
\left [\text{sech}^2 \left (  \frac{ \beta \left  ( r-\mu \right )  }  {2} \right )
+ \text{sech}^2 \left (  \frac{ \beta \left  ( r + \mu \right )  }  {2} \right )  \right ]\nn
& = -\frac{2.16 \,e^2 \,\tau\sqrt{a}} {2\,\hbar^2\, b\,(\pi\,\beta)^{3/2}}
\left[ \text{Li}_{3/2}(-e^{\beta\,\mu} )  +  \text{Li}_{3/2}(-e^{-\beta\,\mu} ) \right],
\end{align}
\begin{align}
\sigma _{ y y}^{\text{dc}}
= \mathcal{L}_{y }^0 & = 
\frac{ e^2\,\tau\, b \,\beta}     { 32\,\pi^2\,\hbar^2\,\sqrt{a}}
\int_0^\infty dr \int_0^{2\,\pi} d\theta  \,\sqrt{r \, |\sec \theta| }\,\sin^2 \theta  
\left [\text{sech}^2 \left (  \frac{ \beta \left  ( r-\mu \right )  }  {2} \right )
+ \text{sech}^2 \left (  \frac{ \beta \left  ( r + \mu \right )  }  {2} \right )  \right ]\nn
& = -\frac{ 3.5\, e^2\,\tau\, b}     {8\,\pi^{3/2} \, \hbar^2\,\sqrt{a\,\beta}}
\left[ \text{Li}_{1/2}(-e^{\beta\,\mu} )  +  \text{Li}_{1/2}(-e^{-\beta\,\mu} ) \right],
\end{align}
\end{widetext}
where $\text{Li}_s(z)$ denotes the  polylogarithm function. 
For $\mu/(k_B\,T)\gg1$, we obtain:
\begin{align}
& \sigma _{x x}^{\text{dc}}= \frac{2.88\,e^2 \,\tau\,\sqrt{a}} 
{2\,\pi^2\,\hbar^2 \,b}
\left[ \mu^{3/2}+\frac{\pi^2}{8\,\sqrt{\mu}}(k_B\,T)^2\right] , \\
& \sigma _{y y}^{\text{dc}}= \frac{7\,e^2\,\tau\, b}     
{ 8\,\pi^2\,\hbar^2\,\sqrt{a}}
\left[\sqrt{\mu}-\frac{\pi^2}
{24\,\mu^{3/2}}(k_B\,T)^2\right].
\label{eq:sigmalowtemp}
\end{align}
Evidently, the low-temperature longitudinal dc conductivities are direction-dependent, and have different doping dependence as well. This is because the group velocity $v_{\bf k}=\left(\frac{1}{\hbar }\frac{\partial \epsilon^s_{\bf k}}{\partial k_{\alpha}}\right)$ in Eq.~(\ref{eq:transcoeff}) differs in the $x$ and $y$ directions as $v_x\sim k_x\, \sigma_x\,,
$ and $ v_y\sim \sigma_y$. This is in contrast to the case of isotropic Dirac Hamiltonian such as graphene, where $v_x\sim\sigma_x$ and $v_y\sim\sigma_y$. Consequently, we obtain $\sigma_{xx}=\sigma_{yy}\sim \mu$, as derived in Appendix~\ref{dirac}. Thus, the anisotropic band spectrum, or in other words, the DOS of the system plays an important role in revealing the anisotropic dc conductivities. We note that for 3D double-Weyl Dirac semimetals with quadratic dispersion in the $xy$-plane $\Bigg[$ with energy spectrum $\epsilon_k=\sqrt{\frac{\hbar ^2(k_x^2+k_y^2)^2}{2 \,m}+v^2 \, k_z^2} \,\Bigg]$, the DOS turns out to be $\rho(\epsilon)\sim |\epsilon|$ similar to 2D graphene. Thus, the $z$-component of the dc conductivities shows dependence similar to that of graphene. However, the $x$ and $y$-components depend quadratically on both chemical potential and temperature \cite{Gegory2016}. But this scenario differs from the 2D model discussed in this paper.    

The thermoelectric coefficients are obtained in a similar fashion, as shown below:
\begin{widetext}
\begin{align}\label{Lxx21}
L_{x x}^{21}&=\frac{e \,\tau\,\sqrt{a}\,\beta} {8\,\pi^2\,\hbar^2\, b}
\int_0^\infty dr \int_0^{2\,\pi} d\theta  \,r^{3/2} \,|\cos \theta|^{5/2} 
\Big [
\mu \left \lbrace  \text{sech}^2 \left (  \frac{ \beta \left  ( r +\mu \right )  }  {2} \right )
+ \text{sech}^2 \left (  \frac{ \beta \left  ( r - \mu \right )  }  {2} \right )  \right \rbrace \nn
& \hspace{ 8 cm} +  r  \left \lbrace  \text{sech}^2 \left (  \frac{ \beta \left  ( r + \mu \right )  }  {2} \right )
- \text{sech}^2 \left (  \frac{ \beta \left  ( r - \mu \right )  }  {2} \right )  \right \rbrace
\Big  ]\nn
& =  -\frac{2.16\, e\, \tau\,\sqrt{a}} {2\hbar^2 \,b\,(\pi \, \beta)^{3/2}}
\left[ 
\mu \left \lbrace \text{Li}_{3/2}(-e^{-\beta\,\mu} )  +  \text{Li}_{3/2}(-e^{\beta\,\mu} )
\right \rbrace
+ \frac{5}{2\,\beta} \left \lbrace \text{Li}_{5/2}(-e^{-\beta\,\mu} )  -  \text{Li}_{5/2}(-e^{\beta\,\mu} )
\right \rbrace \right],
\end{align}
\begin{align}
L_{yy}^{21}&= \frac{ e\,\tau \,b \,\beta}     { 32\,\pi^2\,\hbar^2\,\sqrt{a}}
\int_0^\infty dr \int_0^{2\,\pi} d\theta  \,\sqrt{r \, |\sec \theta| }\,\sin^2 \theta  
\Big [
\mu \left \lbrace  \text{sech}^2 \left (  \frac{ \beta \left  ( r +\mu \right )  }  {2} \right )
+ \text{sech}^2 \left (  \frac{ \beta \left  ( r - \mu \right )  }  {2} \right )  \right \rbrace \nn
& \hspace{ 8.5 cm} +  r  \left \lbrace  \text{sech}^2 \left (  \frac{ \beta \left  ( r + \mu \right )  }  {2} \right )
- \text{sech}^2 \left (  \frac{ \beta \left  ( r - \mu \right )  }  {2} \right )  \right \rbrace
\Big  ]\nn
& = -\frac{ 3.5\, e\,\tau\, b}     { 8\,\pi^{3/2}\,\hbar^2\,\sqrt{a \, \beta}}
\left[ \mu \left \lbrace \text{Li}_{1/2}(-e^{-\beta\,\mu} )  +  \text{Li}_{1/2}(-e^{\beta\,\mu} )
\right \rbrace
+ \frac{3 }{2\,\beta} \left \lbrace \text{Li}_{3/2}(-e^{-\beta\,\mu} )  
-  \text{Li}_{3/2}(-e^{\beta\,\mu} )
\right \rbrace \right] ,
\end{align}
\end{widetext}
At low temperatures, i.e., $\mu/(k_B\,T)\gg 1$, we obtain:
\begin{align}
& L _{x x}^{21}= -\frac{2.88\,e \,\tau\,\sqrt{a}} {2\,\pi^2\,\hbar^2\, b} \times \frac{\pi^2\, \mu^{1/2}}{2}
\left (k_B\,T \right )^2\,,\nn
& L_{y y}^{21}= -\frac{7\, e\,\tau\, b}{ 8\,\pi^2\,\hbar^2\,\sqrt{a}} \times
\frac{\pi^2}{6\,\mu^{1/2}}\left (k_B\,T \right )^2\,.
\end{align}
As before, the low-temperature behavior of the off-diagonal longitudinal thermal coefficients have an interesting direction dependence on the chemical potential. In contrast, for graphene $L_{xx}^{21} = L_{yy}^{21}
=\pi \left (k_B \, T \right )^2$,which are independent of the chemical potential. Although the individual coefficients in the AWF differ from those in graphene, the Mott relation still prevails at low temperature as follows:
\begin{align}\label{eq:lowfieldTh}
S_{xx}=\frac{L_{xx}^{21}}{T\sigma_{xx}^{\rm dc}}\simeq-\frac{\pi^2\, k_B^2\,T}{2\,e\,\mu}\,,\nonumber\\
S_{yy}=\frac{L_{xx}^{21}}{T\sigma_{xx}^{\rm dc}}\simeq-\frac{\pi^2\,k_B^2\,T}{6\,e\,\mu} \,.
\end{align}
Indeed, at low-temperature and for energy-independent scattering, there is no deviation of thermopower from the usual Mott relation. However, different energy-dependent scattering mechanisms may lead to deviation \cite{Sarma2009} from the linear temperature dependent Mott relation as will be evident shortly.

To investigate the electronic contribution to the thermal conductivity  $\kappa$, we next compute:
\begin{widetext}
\begin{align}
L_{x x}^{22 }&= \frac{\mathcal{L}_{x}^2}
{e^2  \, T }\nonumber\\
& = 
\frac{\tau \, \sqrt{a} \, \beta} {8 \, \pi^2\, \hbar^2 \,b \, T}
\int_0^\infty dr 
\int_0^{2\,\pi} d\theta  \,r^{3/2} \,|\cos \theta|^{5/2} 
\Big [
\text{sech}^2 \left (  \frac{ \beta \left  ( r +\mu \right )  }  {2} \right )
 \left  ( r +\mu \right )^2
+ \text{sech}^2 \left (  \frac{ \beta \left  ( r - \mu \right )  }  {2} \right ) \left  ( r - \mu \right )^2 
\Big  ]\nn
& =  -\frac{2.16\tau\sqrt{a}} {2 \, \hbar^2 \,b \,(\pi\beta)^{3/2}\,T}
\Big [ 
\mu^2 \left \lbrace \text{Li}_{3/2}(-e^{-\beta\,\mu} )  +  \text{Li}_{3/2}(-e^{\beta\,\mu} )
\right \rbrace
+ \frac{5 \,\mu}{\beta} \left \lbrace \text{Li}_{5/2}(-e^{-\beta\,\mu} )  -  \text{Li}_{5/2}(-e^{\beta\,\mu} )
\right \rbrace
\nn & \hspace{ 3.7 cm }+ \frac{ 35 }{4\,\beta^2 } \left \lbrace \text{Li}_{7/2}(-e^{-\beta\,\mu} )  
+  \text{Li}_{7/2}(-e^{\beta\,\mu} )
\right \rbrace \Big ]\,,
\end{align}
\begin{align}
L_{yy}^{22}&= \frac{\mathcal{L}_y^2} {e^2\, T} \nonumber\\
&=
\frac{\tau \, b \,\beta}     { 32\,\pi^2 \, \hbar^2\, \sqrt{a}\,T}
\int_0^\infty dr \int_0^{2\,\pi} d\theta  \,\sqrt{r \, |\sec \theta| }\,\sin^2 \theta  
\Big [
\text{sech}^2 \left (  \frac{ \beta \left  ( r +\mu \right )  }  {2} \right ) \left  ( r +\mu \right )^2
+ \text{sech}^2 \left (  \frac{ \beta \left  ( r - \mu \right )  }  {2} \right ) \left  ( r - \mu \right )^2 
\Big  ]\nn
& = -\frac{ 3.5 e^2\tau b}     { 8 \, \pi^{3/2}\, \hbar^2 \, \sqrt{a \, \beta}\,T}
\Big [ 
\mu^2 \left \lbrace \text{Li}_{1/2}(-e^{-\beta\,\mu} )  +  \text{Li}_{1/2}(-e^{\beta\,\mu} )
\right \rbrace
+ \frac{ 3 \,\mu}{\beta} \left \lbrace \text{Li}_{ 3/2}(-e^{-\beta\,\mu} )  
-  \text{Li}_{ 3/2}(-e^{\beta\,\mu} )
\right \rbrace
\nn & \hspace{ 3.7 cm }+ \frac{ 15 }{4\,\beta^2 } 
\left \lbrace \text{Li}_{ 5/2}(-e^{-\beta\,\mu} )  +  \text{Li}_{ 5/2}(-e^{\beta\,\mu} )
\right \rbrace \Big ] .
\end{align}
\end{widetext}
At low temperatures ($\mu/(k_B\,T)\gg 1$), we obtain:
\begin{align}
L _{x x}^{22}
&= \frac{2.88\,\tau \,\sqrt{a}} {2 \,\pi^2\,\hbar^2\, b \, T}
\left [ \frac{\pi^2\, \mu^{3/2}}{3}(k_B\,T)^2
+ \frac{7\,\pi^4}{40\,\mu^{1/2}}(k_B\,T)^4 \right ] ,\nn
 L_{y y}^{22}
 & = \frac{7\,\tau \,b}     { 8\,\pi^2\,\hbar^2\,\sqrt{a}\,T}
\left[ \frac{\pi^2\, \mu^{1/2}}{3}(k_B\,T)^2-\frac{7\,\pi^4}
{120\,\mu^{3/2}}(k_B\,T)^4 \right ].
\label{eq:LLlowtemp}
\end{align}
Together with Eq.~(\ref{eq:LLlowtemp}) and (\ref{eq:sigmalowtemp}), we recover the Wiedemann-Franz law, $L_{\alpha\alpha}^{22}=\frac{\pi^2\, k_B^2\, T}{3\,e^2}\sigma_{\alpha\alpha}^{\rm dc}$, up to leading order in $k_B\,T$.   Finally, using Eq.~(\ref{eq:kappa}), we get:
\begin{align}
& \kappa_{xx}  =  L^{22}_{xx} - \frac{\left ( L^{21}_{xx} \right )^2 }  {T\, \sigma _{x x}^{\text{dc}}  }
\nn &=\frac{2.88\, \tau\sqrt{a}} {2\,\pi^2\,\hbar^2 \,b \, T}
\left[\frac{\pi^2 \,\mu^{3/2}\, (k_B\,T)^2}{3}-\frac{3\, \pi^4\,(k_B\,T)^3}{40\,\mu^{1/2}}\right], \nn
& \kappa_{yy}  =  L^{22}_{yy} - \frac{\left ( L^{21}_{yy} \right )^2 }  {T\, \sigma _{yy}^{\text{dc}}  }
\nn & =\frac{7\,\tau b}     { 8\,\pi^2\,\hbar^2\,\sqrt{a}\,T}
\left[\frac{\pi^2 \mu^{1/2}\, (k_B\,T)^2}{3}-\frac{31\,\pi^4\, (k_B\,T)^3}{360\,\mu^{3/2}}\right].\nn
\end{align}
As expected, the thermal conductivities shows linear dependence on temperature for both the $x$ and $y$-directions. However, their chemical potential dependences differ by $\mu$ as a result of anisotropic dispersion, as discussed before. We note that we have neglected the phonon contribution to the thermal conductivity for simplicity. Strong contributions from phonons may lead to the violation of the Wiedemann-Franz law. 
 

Let us also state our results in the opposite limit of $\mu/(k_B\,T)\ll 1$. In this high temperature limit, we get:
\begin{align}
&  \sigma _{x x}^{\text{dc}}\simeq
\frac{2.16 \,e^2 \,\tau\sqrt{a}} {2\,\hbar^2\, b\,(\pi\,\beta)^{3/2}}
\left( 1.5303 +\frac{0.3801\,\mu^2}{k_B^2 \, T^2}
\right ),\nn
& \sigma _{yy}^{\text{dc}}\simeq
\frac{ 3.5\, e^2\,\tau\, b}     {8\,\pi^{3/2} \,\hbar^2\,\sqrt{a\,\beta}}
\left (1.2098 +  \frac{0.1187\,\mu^2} {k_B^2 \, T^2}   \right),\nn 
& L _{xx}^{21} =-\frac{2.16 \,e^2 \,\tau \, \sqrt{a}} 
{2\,\hbar^2\, b\,(\pi \,\beta)^{3/2}}
\times 2.3\,\mu  
 \,,\nn
& L _{yy}^{21} =-
\frac{ 3.5\, e^2\,\tau\, b}     {8\,\pi^{3/2}\hbar^2\,\sqrt{a\,\beta}}
\times 0.60\,\mu \,,\nn
& L _{x x}^{22} =
\frac{2.16 \,e^2 \,\tau\sqrt{a}\,\beta^2} {2\,T\,\hbar^2\, b\,(\pi\,\beta)^{3/2}}
\left(16.88 + \frac{ 0.6 \, \mu^2}{k_B^2 \, T^2}\right ),\nn
& L_{yy}^{22} =
\frac{ 3.5\, e^2\,\tau\, b\,\beta^2}     {8\,\pi^{3/2}\, \hbar^2\,\sqrt{a\,\beta}}
\left(6.54- \frac{ 0.15 \, \mu^2} {k_B^2 \, T^2}\,
\right ).
\end{align}
It turns out that the prefactors of both Eqs.~(\ref{sigma}) and~(\ref{Lxx21}) give rise to dominant leading order contributions at high temperatures. Thus, both $\sigma$ and $L^{21}$ scale as $T^{3/2}$. Consequently, we obtain thermopower decaying with temperature. We note that the high temperature behavior can be qualitatively understood by rewriting Seebeck coefficient as
$S_{\alpha\alpha}=\langle \epsilon_k\rangle/T-\mu/(e\,T)$, where $\langle \epsilon_k\rangle
= \sum \limits _{s=\pm}\int \frac{d^2 \mathbf{k}}{(2\,\pi )^2}
\, \epsilon_{\bf k}^s \,F^s({\bf k})\, \big / 
\sum \limits _{ \tilde s =\pm}\int \frac{d^2 \mathbf{k'}}{(2\,\pi )^2} 
\, F^{ \tilde s} ({\bf k'}) $, and $F^s({\bf k})
= \tau \left(\epsilon^s_{\bf k}\right)
\frac{\partial f (\epsilon ^s _{\bf k}) }{\partial \epsilon ^s _{\bf k}}
\left(\frac{1}{\hbar }\frac{\partial \epsilon^s_{\bf k}}{\partial k_{\alpha}}\right)^2 $. Neglecting $\langle \epsilon_k\rangle/T$ at $T\rightarrow \infty$, one arrives at $S_{\alpha\alpha}\simeq -\mu/\left({e\,T}\right)$. We note that generically $\langle \epsilon_k\rangle$ may depend on $\mu$ through the Fermi function, which eventually may lead to different prefactors in $S_{xx}$ and $S_{yy}$ in the high temperature limit, as obtained for the present model. It is also worth pointing out that at high temperatures, the Seebeck coefficient can further be related to entropy using the thermodynamic relation between entropy and chemical potential, usually known as Heikes formula \cite{Chaikin1976}. However, this relation turns out to be valid in all temperature ranges, as will be evident shortly. 

Finally, we would like to point out that, for isotropic Dirac dispersion, the leading order scaling of  $\sigma^{\rm dc}_{\alpha \alpha }$ turns out to be $ \propto T$, whereas $L^{21}_{\alpha \alpha}$ scales as $ T^{2}$. This leads to a temperature-independent thermopower in graphene at high temperatures \cite{Sarma2009}.


\section{Diffusive transport due to disorder}
\label{disorder}

We now consider the case of short-range disorder, which is less
realistic for Weyl/Dirac semimetals, because the relatively
poor screening of charged impurities lead to longer-range potentials. Nevertheless, it is useful to investigate the predictions for the thermal properties in this
case for the purposes of comparison. The short-range disorder potential has the following form:
\begin{align}
V(\mathbf r)=V_0\sum_{i}\delta(\mathbf r-\mathbf r_i)\, ,
\end{align} 
where $\mathbf r_i$ denotes position of impurity potential and $V_0$ denotes the strength of the impurity potential. The scattering time for such disorder potential is calculated to be \cite{orignac} 
\begin{align}
\tau_{\rm dis}=\frac{\tau_0(\epsilon)} {1+0.435\cos\theta } \,,
\end{align}
where $\tau_0(\epsilon)=\frac {\hbar}{\pi\, \gamma\,\rho(\epsilon)}$, $\gamma=V_0^2\, n_{\rm imp}$, and $n_{\rm imp}$ is the impurity concentration. Considering this energy dependence of the scattering rate ($\tau\sim \frac{1}{\sqrt{\epsilon}}$), the transport coefficients at low temperatures $\left (\mu/( k_B \,T )\gg 1 \right )$ are found to be:
\begin{align}
& \sigma _{x x}^{\text{dc}}\simeq \frac{2.88\,e^2 \tau\sqrt{a}} {2\,\pi^2\,\hbar^2\, b}\,\mu\,,~~~~\sigma _{y y}^{\text{dc}}\simeq\frac{7\, e^2\,\tau\, b}     { 8\,\pi^2\,\hbar^2\,\sqrt{a}}\,,\nonumber\\
&  L _{x x}^{21}\simeq-\frac{2.88\, e\, \tau\,\sqrt{a}} {2\,\pi^2\,\hbar^2\, b}\,\frac{(\pi\, k_B\, T)^2}{3},
\quad L _{y y}^{21}\simeq -\frac{7 \,e\,\tau\, b}    
 { 8\,\pi^2\,\hbar^2\,\sqrt{a}}\,\mu\,.
\label{eq:diffusive_sc}
\end{align} 
Evidently, the thermopower $S_{xx}$ follows the Mott relation, whereas $S_{yy}$ turns out to be independent of temperature (up to leading order). In contrast, for short-range disorder, the thermopower in graphene is exponentially suppressed at low temperature since $\tau_{\rm dis}\sim 1/\epsilon$ \cite{Sarma2009}.

\section{Transport in presence of charged impurity scatterings}
\label{coulomb}

Presence of charged impurities in a material acts as dopants, thus shifting the Fermi level away from the nodal points. 
The screened Coulomb potential generated by such impurities is given by:
\begin{align}
V(q)=\frac{4\,\pi\, e^2}{q+q_{\text\tiny{TF}}}\,,
\end{align}
where $q_{\rm TF}$ is the Thomas-Fermi wave-vector. The transport relaxation time within the Born approximation is given by:
\begin{align}
& \frac{1}{\tau(\epsilon_{\vec k}^s )}
\nn & = \frac{2\,\pi\,n_{imp} }{\hbar}
\int \frac{d^2 \mathbf{k}'}{(2\,\pi )^2}\, V^2( |\vec k -\vec k'|)\, F_{\bf k\,k'}\,
\delta\big( \epsilon_{\vec k}^s -\epsilon_{\vec k '}^s  \big) \,,
\label{eq:tauscreen}
\end{align}
where $F_{\bf k\,k'}=\frac{1-\cos^2\phi_{\vec k  \vec k'}}{2}$, $\phi_{\vec k  \vec k'}$ is the angle between $\vec k$ and $\vec k'$, and $n_{imp}$ is the impurity density.  
Using the parametrization introduced before, $\cos\phi_{\vec k  \vec k'}$ takes the form:
\begin{align}
& \cos \phi_{\vec k  \vec k'}
\nn & = \frac{s_0 \, \sqrt{\alpha \,|\cos\theta|}\,
\sqrt{\alpha  \, | \cos\theta'|}+ \sqrt{r\,r'}\,\sin\theta\, \sin\theta'
} 
{\sqrt{\alpha\, |\cos\theta|+ r\, \sin^2 \theta } \, \sqrt{   \alpha \, |\cos\theta'| + r'\, \sin^2 \theta' }
}\,,
\end{align}
where $\alpha=b^2/a$, $s_0=\text{sign}[\cos\theta]\,\text{sign}[\cos\theta']$,
and $(r,\,r') \geq 0$.
For definiteness, let us consider the case when $s =+$. 
Since $\epsilon_{\vec k}^+ =r$ is independent of $\theta$, we set $\theta=\frac{\pi}{2}$, without any loss of generality. This leads to 
\begin{align}
F_{\bf k  \, k'}=\frac{\alpha  \, |\cos(\theta')|}
{2 \left ( |\cos\theta'| + r'\sin^2\theta'\right )}\,.
\label{eq:Ftheta}
\end{align}
Together with Eq.~(\ref{eq:Ftheta}), (\ref{eq:tauscreen}), and  (\ref{eq:jacobian}), we obtain 
\begin{align}
 \frac{1}{\tau(r)} 
 &
 = \frac{4\,\pi\, n_{imp}\,e^4\,\alpha} {\hbar\,r^{3/2}}
\int \frac{ d\theta'}
{\left  [
(1-\sin\theta')^2+\frac{\alpha\, |\cos\theta'|}{ r}\right ]}
 \nn & \hspace{ 3. cm} \times  
 \frac{\sqrt{\alpha \,|\cos\theta' |}}  { \alpha\, |\cos \theta '|+r\, \sin^2\theta'  }\,,
 \label{eq:unscreenedimp}
\end{align}
where we have considered $q_{TF}=0$ for unscreened charge impurities. 
In this case, Eq.~(\ref{eq:unscreenedimp}) can be further simplified in the various limits as follows (assuming $\alpha\sim 1$):
\begin{align}
&  \frac{1}{\tau(r )}  
 \simeq \frac{ 4\, \pi \,e^4 \,n_{imp}}{ \hbar}
\begin{cases}
& \frac{8.0}{r}\text{ for }  r\ll 1\,,\\
&\frac{ 6.0476 }{r^{5/3}}
+\frac{16.509}{r^{7/3}}- \frac{10.6889}{r^{3}}
 \text{ for }  r \gg1\, ,
\end{cases}
 \label{eqscat22}
\end{align}
The first limit is found from the leading order contribution of
$2\int_{-\pi/2+r}^{\pi/2-r} \frac{ d\theta'\, \sqrt{|\cos\theta' |} } 
{ \frac{|\cos\theta'|}{r} 
\,|\cos \theta '|  } \,,$ whereas the second limit is found from
the leading order contribution of
$4 \int_{0}^{\pi/2-\left ( \frac{4}{r} \right )^{1/3} } 
\frac{ d\theta'\, \sqrt{ |\cos\theta' |} } 
{\left (1-\sin \theta'\right )^2
	\,r \sin^2\theta'}  $.

We emphasize that the scattering from the unscreened Coulomb interaction in graphene is known to be $\tau\sim\epsilon$ irrespective of the values of $\epsilon$. In contrast, the anisotropy in Eq.~(\ref{ham-continuum}) leads to a different expression for energy-dependent scattering for $\epsilon\gg1$.  Considering the leading energy dependent term for $\tau\sim \epsilon^{5/3}$, we find 

\begin{align}
& \sigma _{x x}^{\text{dc}}= \frac{2.88\,e^2 \,\tau\,\sqrt{a}} 
{2\,\pi^2\,\hbar^2\,  b} \left[ \mu^{19/6}
+\frac{247 \, \mu^{7/6}\pi^2\, (k_B\,T)^2}  { 216} \right] ,\nn
& \sigma _{y y}^{\text{dc}}= \frac{7 \,e^2\,\tau\, b}   
  { 8\,\pi^2\,\hbar^2\,\sqrt{a}}
  \left[ \mu^{13/6}+\frac{91 \, \mu^{1/6}\pi^2 \, (k_B\,T)^2} {216}\right ],\nn
& L _{x x}^{21} =-\frac{2.88\, e\, \tau\,\sqrt{a}} {2\,\pi^2\,\hbar^2\, b}\times
\frac{19\,\pi^2\mu^{13/6}\, (k_B\,T)^2}
{18}
,\nn
& L_{yy}^{21} = -\frac{7 \,e\,\tau\, b}    
{ 8\,\pi^2\,\hbar^2\,\sqrt{a}}\times
\frac{13\,\pi^2\,\mu^{7/6} \, (k_B\,T)^2}
{18}.
\label{eq:diffusive_sc2}
\end{align}

Thus we recover the Mott relation of $S_{\alpha\alpha}\sim T$. However, the dc conductivities have an interesting chemical potential dependence due to energy-dependent scatterings. This is in conjunction with the results obtained before.

\begin{figure}
	\includegraphics[width = 0.99 \linewidth]{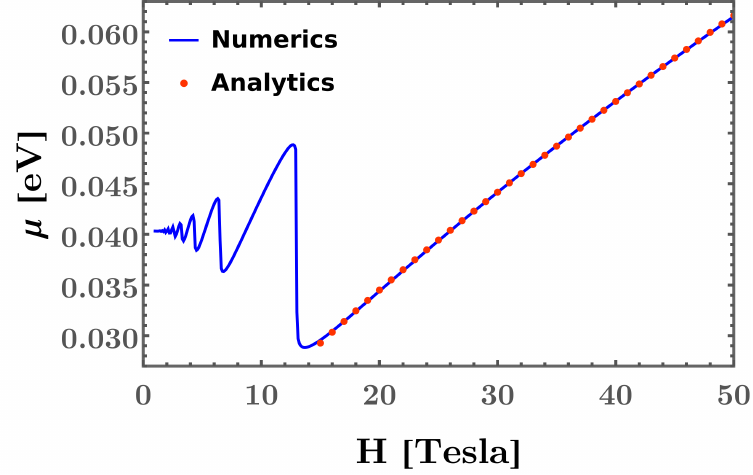}
\caption{Plot (blue solid line) of chemical potential as a function of the magnetic field strength for an electron density of $n_0=5\times 10^{11} \rm{cm}^{-2}$ at temperature $T=5 ~ \rm K$. In the strong field regime, the red dots represent the approximate analytical results of Eq.~(\ref{eq: mufit}), using the values $ b_0=0.0017$ eV, $b_1=0.0006$ eV Tesla$^{-2/3}$, $b_2 = 0.0014 $ eV, and $ b_3 = 1 $ Tesla$^{-1}$ as fitting parameters. Considering typical parameter values for Dirac materials, we have used $v=5\times 10^5 $ m/s and $m=3.1\times m_e$ \cite{Kaplunov,osada}, where $m_e$ is the electron mass. With this mass, the cyclotron frequency $\omega_c$ has a range of $50~ \rm GHz$ to $2~\rm THz$, for the range of magnetic field strength presented in the plot.}
	\label{fig:chempot}
\end{figure}

\section{Thermopower in presence of a quantizing magnetic field}
\label{high magnetic field}

Having obtained the zero magnetic field thermopower, we next turn to the thermopower in the presence of a quantizing magnetic ({\it i.e.}, orbital motion is fully quantized) applied along the $z$-direction, which basically corresponds to the high field limit. In this case, the transverse thermoelectric coefficients $L_{xy}$ and $L_{yx} $ are nonzero. Thus, the Seebeck coefficients are given by [see Eq.~\eqref{eq:general}]:
\begin{align}
S_{xx}=\frac{L^{11}_{yy}\,L^{12}_{xx}+L^{11}_{xy}\,L^{12}_{xy}}
{L^{11}_{yy}\,L^{11}_{xx}+L^{11}_{xy}\,L^{11}_{xy}}\,,\nonumber\\
S_{yy}=\frac{L^{11}_{xx}\,L^{12}_{yy}+L^{11}_{xy}\,L^{12}_{xy}}
{L^{11}_{yy}\,L^{11}_{xx}+L^{11}_{xy}\,L^{11}_{xy}}\,.
\end{align}
Here we have used the fact that $L^{ij}_{yx}=-L^{ij}_{xy}$ in the strong field limit \cite{Girvin_1982}. We now focus on the dissipationless limit, where the heat current is obtained by Hall edge, because of the diverging scattering time. In this limit, $L_{xy}^{11}\gg L_{xx}^{11},\, L_{yy}^{11}$, which in turn leads to  
$S_{xx}\simeq L_{xy}^{12}/L_{xy}^{11}=S_{yy}$. Thus, the Seebeck coefficient in the dissipationless limit turns out to be symmetric in both $x$ and $y$-directions, as opposed to the cases discussed in the preceding sections where $S_{xx}\ne S_{yy}$ (without transverse coefficients). In the following, we concentrate only on the Seebeck coefficient along the $x$-direction, allowing a heat current along the same direction. 

A very useful proposition regarding Seebeck coefficient in materials is that it can be thought of as electronic entropy per unit (net) charge density, i.e., $S_{\alpha\alpha}=\frac {\mathcal S} {e\,n_0}\,$, where $\mathcal S$ is the total entropy, and $n_0$ is the electron density. Although this idea was subject to considerable debates for several years \cite{review}, it is now a well-accepted fact, and there is an extensive literature to support this \cite{Bergman2010,alan, Chaikin1976,yu1965}. 
We note that this relation between thermopower and entropy holds at all temperature, and even in the dissipationless limit. In Appendix~\ref{entropy}, we provide the relation between $L_{xy}$ and the electronic entropy of materials. With this, we proceed to find the thermopower in the presence of magnetic field.     

The total entropy can be expressed in terms of the Fermi-Dirac distribution $f$ function as \cite{Skinner2018,Bergman2010}:
\begin{align}
\mathcal S=-k_B \sum \limits _{n}
\left[ f_n \ln f_n+ \left (1-f_n \right )\ln \left (1-f_n \right ) \right],
\end{align}
where $f_n=f(\epsilon_n-\mu)$, and $\epsilon_n$ denotes the Landau level energy. 
For a magnetic field $\mathbf H = H \,\hat{\mathbf z}$, and using the Landau gauge $\boldsymbol{A} = (- H\,y,0, 0)$, the Landau levels are obtained to be \cite{landau-level}:
\begin{align}
\epsilon_n & = \pm 1.17325
\left (m\, v^2\right )^{1/3}
\left  [ \, \left (n+1/2\right )\hbar \,\omega_c\right ]^{2/3} \,,\nn 
\text{for } & n \in \lbrace 0, \, {\mathbb{Z}^+} \rbrace.
\label{WKB1} 
\end{align}
Here, $\omega_c = \frac{ e\, H} {m}$ is the effective cyclotron frequency. Using the above, we get:
\begin{align}
S_{xx}=\frac{k_B}{2\, \pi\,n_0\,e\, l_b^2}\sum_n \left[\ln(1+e^{\tilde x_n})-\frac{{\tilde x_n}\, e^{\tilde x_n}}{e^{\tilde x_n}+1}\right],
\label{eq:alphaxx}
\end{align} 
where $\tilde x_n=\beta\,(\epsilon_n-\mu)$, $l_b=\sqrt{\frac{\hbar}{e\,H}}$ is the magnetic length, and $n_0$ fixes the Fermi energy through
\begin{align}
n_0=2\times \frac{1}{2\,\pi \,l_b^2}\sum_{n=0}^{\infty} f_{n}\,.
\label{eq:edos}
\end{align} 
Here the factor of $2$ accounts for the hole Landau levels.

\begin{figure}	
	\includegraphics[width= 0.99 \linewidth]{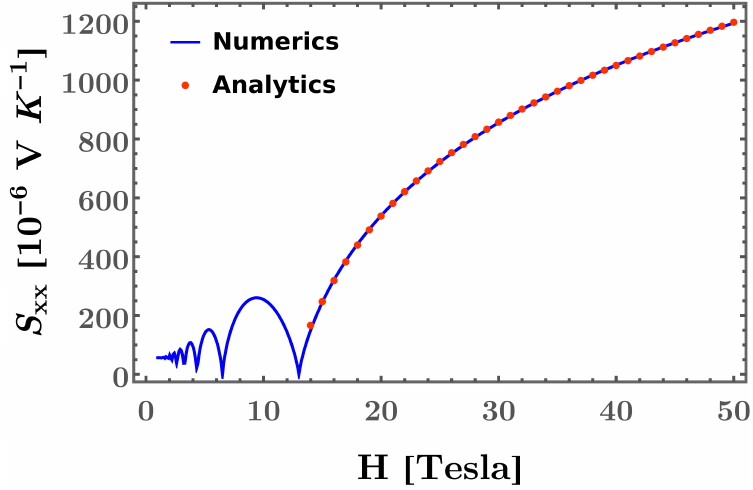}
	\caption{Plot (blue solid line) of $S_{xx}$, as shown in Eq.~\eqref{eq:alphaxx}, as a function of the strength of the magnetic field for an electron density of $n_0 = 5\times10^{11} \,\rm{cm}^{-2}$ at temperature $T = 5\, \rm K$. The red dots represent the approximate analytical results with the function shown in Eq.~(\ref{eq:sxxanaly}). Values of the parameters $v$ and $m$ are the same as in Fig.~(\ref{fig:chempot}).}
	\label{fig:Seebeck}
\end{figure}

For a reasonably strong magnetic field $(\hbar\omega_c\gg\mu)$, the system enters into a strong quantum limit and only the $n=0$ Landau levels are populated. With this assumption, we can approximate Eq.~(\ref{eq:edos}) as:
\begin{align}
n_0\simeq&\frac{1}{\pi \,l_b^2}\times\frac{1} 
{1+e^{\beta \, \left ( |\epsilon_0|-\mu \right )} }\,.
\label{eq:density}
\end{align}
This leads to $\mu= |\epsilon_0| -\beta^{-1}\ln \Big (\frac{1}{n_0 \, \pi\, \l_b^2}-1 \Big)$, which can be expressed in terms of explicit field dependence as
\begin{align}
\mu=b_0 + b_1\,H^{2/3}+b_2\ln \big (b_3 \,H-1 \big ) \,,
\label{eq: mufit}
\end{align}
where the $b_i$'s can be readily obtained from the approximate analytical solution of $\mu$. This approximate analytical result fits reasonably well with the numerical solution obtained from Eq.~(\ref{eq:density}), as corroborated by Fig.~(\ref{fig:chempot}), which justifies its validity. Notably, Eq.~(\ref{eq: mufit}) differs from the case of 3D Dirac/Weyl systems (having $\mu\sim \frac{1}{H}$) and doped semiconductors (having $\mu\sim\frac{1}{H^2}$), as studied in Ref.~\cite{Skinner2018}. This difference again comes from the different magnetic field dependence of the Landau spectrum. Notice that for weak enough magnetic field (characterized by $\hbar \, \omega_c \ll \mu$), the chemical potential is mainly unaffected by the field. As we increase the  field, we start to see quantum oscillations in the chemical potential, which in turn leads to oscillations in the thermopower, as will be evident shortly.

To find the approximate high field dependence of the thermopower, we substitute Eq.~(\ref{eq: mufit}) in Eq.~(\ref{eq:alphaxx}) setting $n=0$. This gives
\begin{align}
S_{xx}&=\frac{k_B}{e}\left[\left(\pi \,n_0 \, l_b^2\right)\ln\left(1-\pi\, n_0\, l_b^2\right)-\ln\left(\frac{1}
{\pi \, n_0\, l_b^2}-1\right)\right]\nonumber\\
&=\frac{k_B}{e}\left[\left(1-\frac{H}{\alpha_0}\right)\ln\left(1-\frac{\alpha_0}{H}\right)-\ln\left(\frac{\alpha_0}{H}\right)\right],
\label{eq:sxxanaly}
\end{align}
where $\alpha_0=\frac{n_0 \,h} {2 \, e}$.  To verify this analytical approximation, we numerically compute Eq.~(\ref{eq:alphaxx}) using the numerical solution of $\mu (H)$. In Fig.~(\ref{fig:Seebeck}), we have plotted the behavior of $S_{xx}$ as a function of $H$. Clearly, the approximate large field dependence of $S_{xx}$ (red dots) fits well with the numerical solutions (solid blue line).
 
Note that, this field dependence differs from the behavior of doped semiconductors and from typical Dirac/Weyl systems \cite{Liang2017}, where $S_{xx}\sim H^2$. We would like to point out that the thermopower here turns out to be large compared to 2D semiconductors such as GaAs/Ga$_{1-x}$, Al$_x$As and Si-metal-oxide-semiconductor field-effect transistors \cite{Fletcher_1999}. This is indeed due to the Dirac nature of the quasiparticles with low Dirac velocity and low zeroth order Landau energy, as pointed out by several authors in the context of graphene \cite{Wei2009,osada}. Interestingly, the thermopower obtained for the present case has good agreement with the experimental results as found in $\alpha$-(BEDT-TTF)$_2$I$_3$ \cite{osada}.  

\section{Thermopower in presence of a weak magnetic field}
\label{weak_field}

In this section, we discuss the form of the Seebeck coefficients at low strengths of a magnetic field applied along the $z$-axis. In the dissipationless limit, the low field behavior can be understood easily, considering temperatures much smaller than the chemical potential. For $k_B \,T\ll \mu$, the entropy can be approximated as $ \mathcal S
\simeq \frac{\pi^2}{3} \, \rho(\mu)\,k_B^2 \,T$ \cite{philip}. At low temperatures and sufficiently weak magnetic fields (i.e., for $k_B \,T\ll \mu \ll \hbar \, \omega_c$), multiple Landau levels are filled. In such a scenario, $\rho(\mu)$ can be approximated by the zero-field DOS given in Eq.~(\ref{eq:rho}), with $\mu\simeq n_0^{2/3} \left (4 \,\pi^2\sqrt{a \,b^2}/3.5 \right )^{2/3}$. With this, we recover the typical temperature and chemical potential dependencies of the thermopower as $S\sim k_B^2 \,T/\mu$.  

The approach used in the preceding section is valid in the strong magnetic field limit, namely $\hbar\,\omega_c\ll\mu$. However, at small magnetic fields, quasiparticle scatterings can be taken into account using the Boltzmann's quasi-classical theory. Within this theory, the thermoelectric coefficients can be expressed as: 
\begin{align}
\label{eq:lowfield}
L_{\alpha\beta}^{11}  & =-\int d\epsilon\, 
\frac{ \partial f(\epsilon)} {   \partial \epsilon} \, \tilde \sigma_{\alpha\beta}(\epsilon)\,, 
\nonumber\\
L_{\alpha\beta}^{12}  & =-\frac{e}{T}\int d\epsilon\,  
\frac{ \partial f(\epsilon)} {  \partial \epsilon}
\, (\epsilon-\mu)\,\tilde \sigma_{\alpha\beta}(\epsilon)\,, 
\end{align} 
where the $\tilde \sigma_{\alpha\beta}$'s are the components of the energy-dependent tensor \cite{girish1,Gegory2014}
\begin{align}
\tilde \sigma (\epsilon)
=
\sigma_0(\epsilon)
\begin{pmatrix} 
v_{x}^2(\epsilon)
&-H\, \tau(\epsilon)\,\tilde v_x\cr H\,\tau(\epsilon) \, \tilde v_y(\epsilon)
&v_y^2(\epsilon)
\end{pmatrix} \,.
\end{align} 
Here,
\begin{align}
& \sigma_0(\epsilon)= e^2\,\rho_0(\epsilon) \,\tau(\epsilon)\,,
\quad \rho_0(\epsilon)=\sqrt{\frac{\epsilon}{4\, a \,b^2}} \,,
\nn &
\tilde v_x (\epsilon)
= v_x^2(\epsilon)\,
{\partial^2 \epsilon \over \partial ^2 k_y}-v_x(\epsilon)\, v_y(\epsilon)\,
 {\partial^2\epsilon \over \partial k_x \, \partial k_y}\,,
\nn & \tilde v_y (\epsilon) =
v_y^2(\epsilon)\,{\partial^2 \epsilon \over \partial ^2 k_x}
- v_x(\epsilon)\, v_y(\epsilon)\,
 {\partial^2 \epsilon \over \partial k_x \, \partial k_y} \,, \nn
 &
 v_x^2(\epsilon) = {2.88\,a\, \epsilon} / {\pi^2} \,, \quad
 v_y^2(\epsilon)=7\,b^2 \,.
\end{align}
We note that the diagonal elements of $ \tilde \sigma(\epsilon)$ are taken up to zeroth-order in $H$ and. for off-diagonal components, we retain terms leading order in $H$. For simplicity, we assume $\tau$ to be independent of the energy. For $k_B \,T\ll \mu$ and $ \hbar \, \omega_c\ll \mu$, Eq.~(\ref{eq:lowfield}) can be further simplified as:
\begin{align}
\label{eq:lowfieldS}
L_{\alpha\beta}^{11}\simeq  \tilde \sigma_{\alpha\beta}(\mu)\,,
\quad 
L_{\alpha\beta}^{12}
\simeq \frac{\pi^2 k_B^2 \, T} {3 \, e} 
\,
\frac{d  }   {d\epsilon}
\tilde \sigma_{\alpha\beta}(\epsilon)
\Bigg |_{\epsilon=\mu}\,. 
\end{align} 
With this, we obtain the thermopower coefficients as:
\begin{align}
S_{xx}
& \simeq \frac{\pi^2\, k_B^2\, T}{3\,e\,\mu} 
\nn &  \times
 \frac{1.5\, v^2_x(\epsilon)\,v^2_y(\epsilon)
 + H^2\,\tau^2\, \tilde v_x(\epsilon)
\left[0.5\,\tilde v_{y}(\epsilon) + \mu\,
\partial_\epsilon \tilde v_{y} (\epsilon)\right]}
{v^2_x(\epsilon)\,v^2_y(\epsilon)
+ H^2\,\tau^2\, \tilde v^2_x(\epsilon)\,
 \tilde v^2_y(\epsilon)} 
\Bigg |_{\epsilon=\mu} \,,
\nonumber\\
S_{yy}
& \simeq 
\frac{\pi^2\, k_B^2 \,T}{3\,e}\\
&\times \frac{ \frac{0.5\, v^2_x(\epsilon)\,v^2_y(\epsilon)} {\mu}
+H^2\,\tau^2\,\tilde v_y(\epsilon)
\left[0.5\,\tilde v_{x}(\epsilon)+\mu\, \partial_\epsilon {\tilde v_{x}} (\epsilon)\right]}
{v^2_x(\epsilon)\,v^2_y(\epsilon)+H^2\,\tau^2\,\tilde v^2_x(\epsilon)\,\tilde v^2_y(\epsilon)}
\Bigg |_{\epsilon=\mu} \,.
\end{align}
Evidently, in the limit of $H\rightarrow 0$, we recover the field-free theromopower as shown in Eq.~(\ref{eq:lowfieldTh}).

Finally, we comment on the behaviour of the transverse thermoelectric coefficient $S_{xy}$ (or $S_{yx}$), also dubbed as the magneto-thermoelectric Nernst-Ettinghausen effect. For simplicity, we focus on $S_{xy}$ [which is given by Eq.~(\ref{eq:general})]:
\begin{align}
S_{xy} = \frac{L^{12}_{xy} \, L^{11}_{yy}-L^{12}_{yy} 
L^{11}_{xy}} {L^{11}_{yy} \, L^{11}_{xx}-L^{11}_{xy} \,L^{11}_{yx}}
\,.
\end{align}
Clearly, for zero values of the transverse coefficients $L^{11}_{xy}$ and $L^{12}_{xy}$, $S_{xy}$ is identically zero. In the weak field limit when $L_{xx}^{11}\gg L_{xy}^{11}$, $S_{xy}$ is found to be $S_{xy}\sim {L_{xy}^{12}}  / {L_{xx}^{11}}$, which in turn leads to the usual Mott relation [as is evident from Eq.~(\ref{eq:lowfield})]. We note that $S_{yx}$ also follows Mott relation but the prefactor differs from $S_{xy}$ due to anisotropy. In contrast, for strong magnetic field, the two terms in the numerator of $S_{xy}$ mutually cancel/reduce each other since $L_{yy}^{11}\ll L_{xy}^{11}$ and/or $L_{yy}^{12}\ll L_{xy}^{12}$ (dissipationless limit), similar to the results obtained in Refs.~\cite{Girvin_1982,Checkelsky2009}.

\section{Conclusion}
\label{conclusion}
In this paper, we have studied the zero and finite magnetic field thermoelectric coefficients in an anisotropic 2D Weyl system, with the two anisotropic directions having linear and quadratic dispersions respectively. We have shown that this intrinsic anisotropy leads to an interesting doping and temperature dependence of the thermopower, compared to its isotropic counterpart. Our findings can be summarized as follows:
\rm{(i)} The low temperature dc conductivities have a different Fermi energy dependence than the case of graphene (with 2D isotropic Weyl dispersion).
\rm{(ii)} the high temperature thermopower decays with temperature in AWF, whereas it is independent of temperature in graphene.
\rm{(iii)} The relaxation rates due to diffusive and electron-electron interactions differ from the case of graphene, resulting in distinct expressions for the thermal and dc conductivities.
\rm{(iv)} The finite field thermopower has an interesting magnetic field dependence, resulting in unsaturated thermopower. We note that the results obtained here for a single node anisotropic Dirac/Weyl system can be used for multinode systems, provided that there is no internode scattering.  

We conclude that the doping and temperature dependence of the transport measurements can be used to distinguish Dirac materials exhibiting anisotropy. In addition, the field-dependent large thermopower can have potentials for thermoelectric devices to transform heat into electric power. In future work, it will be worthwhile to analyze the effects of Coulomb as well as short-range four-fermion interactions, and impurities, as has been done in the case of 2D \cite{kai-sun,ips-sebgem} and 3D \cite{rahul-sid,ips-rahul,ips-qbt-sc} isotropic semimetals with quadratic band touching points.


\begin{widetext}
	\appendix
	
	\section{Thermoelectric response for the 2D Weyl semimetal}
	\label{dirac}
	
In this appendix, we compute the response matrix for the 2D isotropic Weyl semimetal, with the Hamiltonian 
	\begin{align}
	\label{diracham}
	H_{D} = v \left (  k_x  \, \sigma_x + k_y   \,\sigma_y \right ) .
	\end{align}
	Here we can use the usual polar coordinate parametrization $ k_x= r\,\cos \theta $ and $ k_y= r  \, \sin \theta$ with $ r \geq 0$, such that the energy eigenvalues are given by $ \epsilon_{\bf k}^\pm = \pm\,v\, r$.
The density of states is $\rho(\epsilon) =\frac{|\epsilon|}{2\pi\,v^2}\,$.

We compute the dc conductivity by assuming an energy and momentum independent scattering time, such that
	\begin{align}
	 \sigma_{x x}^{\text{dc}} =\sigma _{ yy }^{\text{dc}} 
	= \mathcal{L}_{x }^0 = \mathcal{L}_{ y }^0 
	&    = 
	\frac{\beta\,v^2\,e^2 \,\tau  }     { 8 \,\pi\, \hbar^2}
	\int_0^\infty dr   \,r  
	\left [
	\text{sech}^2 \left(  \frac{\beta  \left(r+ \mu \right ) } {2} \right )
	+ \text{sech}^2 \left(  \frac{\beta  \left(r -\mu \right ) } {2} \right ) \right ]
	 = \frac{e^2\, \tau \,\ln \left[ 2+ 2\,\cosh \left ( \beta \,\mu \right ) \right ]}
	{4\, \pi\, \hbar^2\,\beta }\,.
	\end{align}
At low temperatures ($\mu/(k_B T)\gg 1)$, this yields $ \sigma_{x x}^{\text{dc}} \sim\mu$.

	The thermoelectric coefficients are given by:
\begin{align}
	& L_{x x}^{21}= {L}_{ yy }^{21}   = \frac{-\mathcal{L}_{x }^1}{e} = \frac{-\mathcal{L}_{y }^1}{e} 
	\nn &= 
	\frac{\beta\,v^2\,e  \,\tau}     {8\,\pi^2\,  \hbar^2}
	\int_0^\infty dr \,r
	\left [
	\mu \left \lbrace  \text{sech}^2 \left (  \frac{ \beta \left  ( r +\mu \right )  }  {2} \right )
	+ \text{sech}^2 \left (  \frac{ \beta \left  ( r - \mu \right )  }  {2} \right )  \right \rbrace 
	+  r\,  \left \lbrace  \text{sech}^2 \left (  \frac{ \beta \left  ( r+ \mu \right )  }  {2} \right )
	- \text{sech}^2 \left (  \frac{ \beta \left  ( r- \mu \right )  }  {2} \right )  \right \rbrace
	\right  ]\nn
	& =  -\frac{v \,e \,\tau}
	{ \left(2\,\beta \, \pi \, \hbar  \right )^2 }
	\left[ 
	\beta \,\mu   \,\ln \left \lbrace  2+ 2\,\cosh \left ( \beta \,\mu \right ) \right  \rbrace 
	+ 2 \, {\text{Li}}_{2} (-e^{ \beta\,\mu} )  - 2 \, {\text{Li}}_{2}(-e^{ -\beta \,\mu} ) 
	\right],
\end{align}

	\begin{align}
	& L_{x x}^{22 } = L_{ yy }^{22 } = \frac{\mathcal{L}_{x }^2}{e^2  \, T } 
	= \frac{\mathcal{L}_{ y }^2}{e^2  \, T } 
	\nn &= 
	\frac{\beta\,v^2\,\tau}     {8\,\pi\,  \hbar^2\, T}
	\int_0^\infty dr  \,r\,
	\left [
	\text{sech}^2 \left (  \frac{ \beta \left  ( r +\mu \right )  }  {2} \right ) \left  ( r\,\epsilon_0 +\mu \right )^2
	+ \text{sech}^2 \left (  \frac{ \beta \left  ( r- \mu \right )  }  {2} \right ) \left  (r-\mu \right )^2 
	\right  ]\nn
	& =  \frac{v \,\tau}
	{4\,\pi\,\hbar^2\, T }
	\left [  
	\frac{4\,\mu 
		\left \lbrace {\text{Li}}_{2} (-e^{- \beta\,\mu} )  - {\text{Li}}_{2}(-e^{ \beta \,\mu} )  \right  \rbrace} {\beta} 
	+ \frac{   6 \, {\text{Li}}_{3} (-e^{ \beta\,\mu} ) +  6 \, {\text{Li}}_{3} (-e^{ -\beta\,\mu} ) }   {\beta^2 }
	- \mu^2  \,\ln \left \lbrace  2+ 2\,\cosh \left ( \beta \,\mu \right ) \right  \rbrace
	\right ] .
	\end{align}
At low temperatures, we get:
\begin{align}
L_{xx}^{21}=L_{yy}^{21}=\frac{v^2\,e\,\tau}
{2\,\pi\,\hbar^2}\times \frac{\pi^2 \,(k_B \,T)^2}{3}
\,, \quad
L_{xx}^{22}=L_{yy}^{22}=
\frac{v^2\,\tau}{2\,\pi\,\hbar^2}\times \frac{\mu\,\pi^2 \,k_B^2\, T}{3}\,.
\end{align}

\section{Relation between the Seebeck coefficient and entropy}
\label{entropy}	

In order to derive the relation between entropy and the Seebeck coefficient, in the presence of sufficiently strong magnetic fields, we begin with the general expression of thermoelectric coefficients $L^{12}_{xy}$ and $L^{11}_{xy}$ \cite{Girvin_1982}:
\begin{align}
\label{eq:Hall}
L^{11}_{xy}=-\frac{e^2}{h}\sum_{n}\int_{\epsilon_n-\mu}^{\infty}d\epsilon\, {\partial f(\epsilon)\over\partial \epsilon}\,,
\quad 
L^{12}_{xy}
= \frac{k_B\,e\,\beta}{h}\sum_{n}\int_{\epsilon_n-\mu}^{\infty}d\epsilon\, \epsilon\,{\partial f(\epsilon)\over\partial \epsilon}\,,
\end{align} 
where $\epsilon_{n}$ denotes the Landau energy spectrum and $f(\epsilon)=\frac{1}{1+e^{\beta \epsilon}}$. Note that the transport properties are independent of the details of the confining potential of the sample, although microscopic currents depend on it. Eq.~(\ref{eq:Hall}) can further be simplified by changing variables $\epsilon\rightarrow f$ as follows:
\begin{align}
L^{11}_{xy}  =-\frac{e^2}{h}\sum_{n}f_n\,,\quad
L^{12}_{xy}
=\frac{k_B\,e}{h}\sum_{n}\int_{\epsilon_n-\mu}^{\infty}\,df\, [\ln (1-f)-\ln f]
=\frac{e}{h} \, \mathcal S \,,
\end{align}  	
where $f_n=f(\epsilon_n-\mu)$, and 
\begin{align}
\mathcal S=-k_B\sum_n\left [ f_n \ln f_n+(1-f_n)\ln (1-f_n) \right ]
\end{align} 
is the total entropy of the carriers. Using this in $S_{xx}=\frac{ \mathcal S}{e\,n_0}$, we obtain the thermopower, where $n_0 $ is the total number of carriers given by Eq.~\eqref{eq:edos}.

\end{widetext}

\bibliography{biblio.bib}

\end{document}